\documentclass[12pt,preprint]{aastex}
\usepackage{pdflscape}

\def\lea{\mathrel{<\kern-1.0em\lower0.9ex\hbox{$\sim$}}}
\def\gea{\mathrel{>\kern-1.0em\lower0.9ex\hbox{$\sim$}}}
\newcommand{\lta}{{\>\rlap{\raise2pt\hbox{$<$}}\lower3pt\hbox{$\sim$}\>}}
\newcommand{\gta}{{\>\rlap{\raise2pt\hbox{$>$}}\lower3pt\hbox{$\sim$}\>}}

\author{Rupali Chandar,\altaffilmark{1} 
S.\ Michael Fall,\altaffilmark{2} and
Bradley C.\ Whitmore\altaffilmark{2}
}

\altaffiltext{1}{Department of Physics \& Astronomy, The University of Toledo, Toledo, OH 43606}
\altaffiltext{2}{Space Telescope Science Institute, Baltimore, MD, USA}

\begin{document}


\title{The Link Between the Formation Rates of Clusters and Stars in Galaxies}

\begin{abstract}

The goal of this paper is to test whether the formation rate of  star clusters is proportional to the star formation rate (SFR) in galaxies. As a first step, we present the mass functions of compact clusters younger than 10~Myr in seven star-forming galaxies of diverse masses, sizes, and morphologies: the Large and Small Magellanic Clouds, NGC~4214, NGC~4449, M83, M51, and the Antennae. These cluster mass functions (CMFs) are well represented by power laws, $dN/dM \propto M^{\beta}$, with similar exponents $\beta = -1.92 \pm0.27$, but with amplitudes that differ by factors up to $\sim$$10^3$, corresponding to vast differences in the sizes of the cluster populations in these galaxies. We then normalize these CMFs by the SFRs in the galaxies, derived from dust-corrected H$\alpha$ luminosities,  and find that the spread in the amplitudes collapses, with a remaining rms deviation of only $\sigma(\log A) = 0.2$. This is close to the expected dispersion from random uncertainties in the CMFs and SFRs. Thus, the data presented here are consistent with exact proportionality between the formation rates of 
stars and clusters. However, the data also permit weak deviations from proportionality, at the factor of two level, within the statistical uncertainties. We find the same spread in amplitudes when we normalize the mass functions of much older clusters, with ages in the range 100 to 400~Myr, by the current SFR.  This is another indication of the general similarity among the cluster populations of different galaxies.
\end{abstract}


\section{Introduction}
Stars form together in clusters and associations, which in turn form in the densest parts---the clumps---of molecular clouds (Lada \& Lada 2003; McKee \& Ostriker 2007). There is some debate about the exact definition of a cluster, e.g., whether this term refers to all dense stellar aggregates or only those that are gravitationally bound or free of gas. Nevertheless, there appears to be unanimous agreement that stars mostly form in aggregates of some sort and rarely in isolation. This general picture implies that the formation rates of stars and clusters should be proportional to each other.
The purpose of this paper is to present a direct test of this proportionality.

Our approach is to compare the mass functions, $\psi(M) = dN/dM$, of recently formed star clusters (with ages $\tau < 10$~Myr) in different galaxies before and after normalizing by the star formation rates in the galaxies.
This is valid because the mass functions of young clusters in different galaxies have similar power-law shapes, $\psi(M) \propto M^{\beta}$ with $\beta \approx-2$, over large intervals of mass, roughly $10^3~M_{\odot} \lea M \lea 10^6~M_{\odot}$ (Fall \& Chandar 2012 and references therein). The mass functions are unknown at lower masses, as a result of selection limits in the available samples, and are uncertain at higher masses as a result of small number statistics. This is the reason that we compare the mass functions,
rather than the total numbers or total masses of  young clusters (the integrals of $\psi(M)$ and  $M \psi(M)$  respectively over all $M$), with the star formation rates.

In this work, we use the term ``cluster'' for any concentrated aggregate of stars with a density much higher than that of the surrounding stellar field, whether or not it is gravitationally bound. It is not possible to tell from images alone which clusters are gravitationally bound (have negative energy) and which are unbound (have positive energy), and N-body simulations show that unbound clusters retain the appearance of bound clusters for many ($\sim$10) crossing times (Baumgardt \& Kroupa 2007).  While spectra might help in principle, they are seldom available and are often contaminated by non-virial motions (stellar winds, binary stars, etc).

In the next section (Section~2), we present the catalogs and  mass functions
in seven galaxies with well-studied populations of clusters. In the following section (Section~3), we estimate the star formation rates, determine the CMF/SFR relations, present a method for comparing this relation between galaxies, and compare the residuals with various galaxy properties. In the last section (Section~4), we summarize some implications of our results for the formation and disruption of the clusters.

\section{Cluster Catalogs and Mass Functions}

Our study focuses on seven nearby galaxies: LMC, SMC, NGC~4214, NGC~4449, M83, M51, and the Antennae. In this section, we collect and describe
the star cluster catalogs, and summarize the method used to estimate the masses and ages of the clusters. We then present the mass functions of the clusters in each galaxy.

\subsection{Star Cluster Samples}

We include nearby galaxies with published star cluster catalogs that cover at least 50\% of the optically luminous portions of the galaxy and have
photometry in at least four optical passbands including the $U$~band
(which is critical for disentangling the effects of age and extinction in the colors of young clusters). Our  sample includes galaxies of different types:
irregular, spiral, and merger, which span a large range in distance, luminosity, color, inclination, and star formation rate. This sample, while small,  is reasonably representative of nearby star-forming galaxies in general.
The observations, data, and cluster selection criteria for each galaxy are summarized below.

{\bf Large and Small Magellanic Clouds:}
We use the catalogs published by Hunter et~al.\ (2003), where clusters were selected by visual examination of candidates (from previously published catalogs). The clusters were distinguished from the surrounding distribution of field stars, resolved with respect to a single star, but do not include low density H\,\textsc{ii} regions. The observations cover approximately 70\% and 90\% of the recent star formation in each galaxy, as traced by H$\alpha$ emission. This resulted in a total of 854 (239) clusters in the LMC (SMC)
with published \textit{UBVR} photometry.

{\bf NGC~4214:} This dwarf irregular galaxy, at a distance of 3.1 Mpc 
(Dopita et~al.\ 2010), was observed in a single pointing with the WFC3 on-board the \textit{Hubble Space Telescope} (\textit{HST}) as part of program 11360 (PI: O'Connell). Clusters were automatically selected to be compact sources broader than the PSF followed by elimination of close pairs of stars, which resulted in a total of 334 candidate clusters. Aperture photometry was performed on \textit{UBVI}H$\alpha$ images in a manner similar to that described in Chandar et~al.\ (2010) for sources in the nearby spiral galaxy M83.

{\bf NGC~4449:}  Rangelov et~al.\ (2011) published a catalog of 129 clusters in this dwarf starburst galaxy (located at a distance of  $3.82\pm0.27$~Mpc; Annibali et~al.\ 2008) based on \textit{HST} observations. They selected cluster candidates automatically using size criteria, followed by a visual inspection to eliminate contaminants such as background galaxies and chance superpositions. Two fields within NGC~4449  were observed with the ACS/WFC in the \textit{BVI}H$\alpha$ bands, and two in the $U$~filter with the WFPC2 camera, giving near-complete coverage of the optically luminous portions of the galaxy.

{\bf M83:}
Two fields within M83, a grand-design spiral galaxy located at a distance of 4.5~Mpc (Thim et~al.\ 2003), were observed as part of program GO-11360.
Several cluster catalogs were produced for each field, according to the selection methodology described in Chandar et~al. (2014).  Here, we adopt the manual cluster catalogs from that work. We also use data from five additional fields in M83 that were observed with the \textit{HST}/WFC3 as part of program GO-12513, bringing the coverage of the optically luminous portion of M83 to $\approx$60\%. New cluster catalogs have been produced for these fields, and will be presented in Whitmore et~al.\ (in prep). Our final catalog contains 3186 star clusters.

{\bf M51:} Most of the optically luminous portions of this grand-design spiral galaxy, located at a distance of  8.2~Mpc, were observed in six fields taken
with the \textit{HST}/ACS camera in \textit{BVI}H$\alpha$, and in 8 overlapping fields with the WFPC2 camera in the $U$~band. Just as for NGC~4449, star clusters were selected using automated criteria, followed by visual examination. This resulted in a catalog of 3812 clusters (Chandar
et~al., in prep).

{\bf Antennae:}
The merging Antennae galaxies, located at a distance of 21~Mpc (Schweizer
et~al.\ 2008), were observed with the ACS/WFC on the \textit{HST} as part of  program GO-10188, in the \textit{BVI}H$\alpha$ filters (Whitmore et~al.\ 2010). Additional observations were taken in the $U$~band with the WFC3 camera as part of program GO-11577. Clusters were selected to be objects brighter than $M_V=-9$, a restriction that eliminates practically all individual stars.

The largest source of uncertainty in these catalogs is in the selection of clusters with ages $\tau\lea10$~Myr. For the Magellanic Clouds, it is possible that clusters that are both very young and/or have low stellar density
were excluded by the selection criteria. For the more distant galaxies, very young clusters might be excluded because they are very compact and their profiles are indistinguishable from individual stars, but on the other hand,
chance superpositions of unassociated young stars might sometimes be selected as young clusters. The selection of compact  clusters older than 
10~Myr is more robust (as discussed in Chandar et~al.\ 2014), because by this age clusters have typically moved away from their crowded birth sites.

\subsection{Mass and Age determinations}

For each cluster in each galaxy, the measured magnitudes, in the filters listed above, were compared with predictions from the Bruzual \& Charlot (2003, 2010) stellar population models, in order to estimate their age, extinction, and mass.

We first estimate the age $\tau$ and the extinction $A_V$ by performing a
$\chi^2$ fit comparing observed and predicted magnitudes, assuming a Chabrier (2003) IMF, and a Galactic-type extinction law (Fitzpatrick 1999).
The best-fit values of $\tau$ and $A_V$ are those that minimize:
\begin{equation}
\chi^2(\tau,A_V) = \sum_{\lambda}
W_{\lambda}~\left(m_{\lambda}^{\mbox{obs}}
- m_{\lambda}^{\mbox{mod}}\right)^2~~,
\end{equation}
where $m_{\lambda}^{\mbox{obs}}$ and $m_{\lambda}^{\mbox{mod}}$ are
the observed and model magnitudes respectively, and the sum runs over all filters listed above. The weight factors in the formula for $\chi^2$ are taken to be $W_{\lambda} = [\sigma_{\lambda}^2 + (0.05)^2]^{-1}$, where $\sigma_{\lambda}$ is the photometric uncertainty.

The mass of each cluster is estimated from the extinction-corrected 
$V$-band magnitude, the distance listed in Table~1, and the mass-to-light
ratio ($M/L_V$) predicted by the models at the fitted ages $\tau$. The models assume that the stellar IMF for each cluster is fully sampled. While the colors of clusters with masses below $\approx$10$^4~M_{\odot}$ begin
to spread around the mean values because the upper end of the stellar IMF
is not fully populated, the resulting mass distributions, down to at least 
$\approx$$3\times 10^3~M_{\odot}$, are very similar regardless of whether fully or stochastically sampled models are used (Fouesneau et~al.\ 2012).

The largest uncertainties in the mass estimates comes from uncertainties in the age estimates, which are typically on the order of 0.3 in $\log\tau$, corresponding to a factor of $\approx$2 in $\tau$. These translate to approximately similar uncertainties of  0.3 and $\approx$2 in $\log M$ and 
$M$. There are also two sources of systematic uncertainty. Any error in a galactic distance translates directly into a fractional change in the masses of all clusters within that galaxy. The derived masses of the clusters, but not their ages, also depend on the IMF assumed in the stellar population models.
For example, if we had assumed a Salpeter (1955) instead of Chabrier IMF, the estimated mass of each cluster would increase by a nearly constant 
$\approx$40\%. Similarly, we find that a mix of stellar IMFs between Chabrier and Salpeter would increase the estimated masses by a nearly constant fractional amount between 0 and 40\% (depending on the exact mix of IMFs). Such fractional changes in $M$, corresponding to shifts in 
$\log M$, would not alter the shape of the cluster mass function (in 
$\log$--$\log$ plots).

\subsection{Cluster Mass Functions}

In the left-hand panels of Figures~1 and 2, we show the mass functions determined for the clusters in all seven of our sample galaxies in two different intervals of age: recently formed ($\tau < 10$~Myr) clusters, which are the main focus of this paper, and also intermediate-age ($\tau=100$--400~Myr) clusters as an interesting check. An equal number of clusters have been placed 
in each bin, from 3 to 7, depending on the available sample size (resulting in variable bin widths). The low end of the mass function in each galaxy is set by the luminosity limit of each catalog, while the high end of these distributions 
is not constrained. The amplitudes of the cluster mass functions directly reflect differences in the sizes of the cluster populations among the galaxies.
There is clearly a large range, approximately a factor of $10^3$, in these amplitudes.

The merging Antennae system has the highest number of stellar clusters
in our sample (in both age intervals). It is followed by the spirals M51 and M83, then the luminous star-forming irregular galaxies NGC~4449 and the LMC, with the low luminosity irregular galaxies NGC~4214 and the SMC
having the fewest detected clusters. The high end of the cluster mass functions also generally correlate with the total number of clusters, in the sense that the galaxies with the most clusters tend to have the most massive
clusters.

The observed mass functions are well represented by featureless power laws: 
$dN/dM \propto M^{\beta}$. Thus, we perform linear fits to 
$\log (dN/d\log M) = (\beta+1)\log M +$~const. The best-fit exponents and their formal $1\sigma$ errors are listed in Table~1. Despite the very different types of galaxies included in our sample, different cluster selection criteria, filter sets, etc.,  the resulting exponents have a fairly narrow range, 
$\beta = -1.92\pm0.27$, indicating that the mass functions of young star clusters in nearby star-forming galaxies have an approximately ``universal" shape.

Uncertainties in the CMF are dominated by the selection of the clusters.
For clusters younger than 10~Myr,  we estimate that the uncertainties in the amplitude of the CMF could be as large as $\approx$0.1 based on comparisons of the cluster catalogs from different groups, which differ at the 
30--40\% level in this age range (e.g., Bastian et~al.\ 2012; Chandar et~al.\  2014). The uncertainty for 100--400~Myr clusters is smaller, roughly 
$\approx$0.04 because cluster catalogs in this age range only differ by 
$\approx$10\% between groups.

\section{Star Formation Rates and Normalized Mass Functions}

\subsection{Star Formation Rates}

We determine the star formation rate for each galaxy in our sample from its 
H$\alpha$ line emission. Most of this emission ($\sim$90\%) is contributed 
by stars younger than 10~Myr (Kennicutt \& Evans 2014). Therefore, 
H$\alpha$ emission measures the SFR over the same age range as the clusters in our younger subsample. We use the most recent H$\alpha$ flux measurements listed in NED and corrected for contamination by 
[N\,\textsc{ii}] emission, and assume the distances given in column~3 of Table~1 when converting to H$\alpha$ luminosities.

The H$\alpha$ emission is attenuated by dust. To correct for this attenuation, we use the measured 25~$\mu$m flux (available for all galaxies in our sample), and the recipe given in Table~2 in  Kennicutt \& Evans (2014):
\begin{equation}
L(H\alpha)_{\mbox{corr}} = L(\mathrm{H}\alpha)_{\mbox{obs}} + 0.020 L(25\mu m)~~.
\end{equation}
We then adopt their Equation~(12) and Table~1 to determine the star formation rate:
\begin{equation}
\log\mathrm{SFR} (M_{\odot}/\mbox{yr}) = 
\log L(\mathrm{H}\alpha)_{\mbox{corr}} - 41.27~~.
\end{equation}
This calibration is based on the STARBURST99 models (Leitherer et~al.\  1999) assuming solar metallicity, a Kroupa IMF (similar to the Chabrier IMF in the Bruzual \& Charlot models), and a constant rate of star formation. The resulting SFRs are compiled in column~5 of Table~1.

There are several inherent sources of uncertainty in H$\alpha$-based star formation rate determinations, including flux measurements, corrections made for attenuation by dust, variations in the star formation rate, uncertainties in the calibration used to convert the measured fluxes to SFR
as discussed in the review by Kennicutt \& Evans (2014), and the leakage of Lyman continuum photons from the parent galaxy. Estimates of these uncertainties are listed in Table~2. For our sample specifically, we find that the published H$\alpha$ and 25~$\mu$m flux measurements (compiled in NED) used to determine the SFRs, typically differ by $\approx$20\%, although one galaxy has published values that differ by $\approx$50\%.
Different methods to correct for dust extinction introduce uncertainties of 
$\approx$20--30\%, based on a comparison between a linear combination of the H$\alpha$ and 24~$\mu$m luminosities with Balmer-decrement corrected H$\alpha$ luminosities (Figure~3, Kennicutt \& Evans 2014), although the uncertainties can be higher for dusty galaxies like the Antennae.
We estimate that the uncertainty in the fractional coverage of each galaxy is 
$\approx$10\%. Lyman continuum photons can escape from their parent galaxy, further increasing the uncertainty in the SFR determinations,
although the escape fraction can vary strongly from one galaxy to another and is therefore difficult to quantify. Taken together, the random uncertainties related to the SFR determinations and coverage correspond to at least 
$\approx$0.2 in the logarithm of the CMF/SFR relation, not including uncertainties related to the cluster mass functions. Systematic uncertainties may also be present, since stellar population modeling shows that the calibrations likely overestimate the SFR for lower metallicity populations, because the ionizing luminosity increases by $\sim$$0.4\pm0.1$~dex for a tenfold decrease in the metallicity  (e.g., Kennicutt \& Evans 2014; Smith, Norris, \& Crowther 2002; Raiter, Schaerer \& Fosbury 2010).

Our H$\alpha$-based star formation rates agree reasonably well with previous
determinations published in the literature (compiled in Column~6 of 
Table~1). Lee et~al.\ (2009) determined H$\alpha$-based SFRs for 
NGC~4214, NGC~4449, M83, and M51, but used a different method to correct for dust and a different calibration to convert H$\alpha$ luminosity to SFR (taken from Kennicutt 1998). James et~al.\ (2008) estimated 
H$\alpha$-based SFRs for the Large and Small Magellanic Cloud using the prescription given by Kennicutt (1998), and Zhang et~al.\ (2001) measured the H$\alpha$ line intensity from \textit{HST}/WFPC2 images and applied the prescription from Kennicutt (1998) to determine the SFR of the Antennae. In Column~7 of Table~1, we list previously published star formation rates determined for each galaxy using various 
non-H$\alpha$-based methods. Overall, we find that our SFR determinations
agree with published values to somewhat better than a factor of two, which is consistent with the uncertainties found by Lee et~al.\ (2009) by comparing far ultraviolet and H$\alpha$-based SFRs for a much larger sample of galaxies.

\subsection{The CMF/SFR Relation}

The right panel of Figure~1 presents the main result of this paper. The same cluster mass functions as in the left panel are shown, but these distributions have now been divided by the star formation rate (listed in Column~5 of Table~1) of the host galaxy within the area covered by each cluster catalog:
CMF/SFR. {\em The mass functions of star clusters in these very different galaxies all have very similar amplitudes when normalized by the star formation rate of their host galaxies.}

In Figure~2, we again show the observed and SFR-normalized cluster mass functions, but now for clusters with ages between 100 and 400~Myr.
Evidently, the mass functions of these intermediate-age clusters also lie very close to one another in the vertical direction after they have been normalized by the current star formation rate of the host galaxy. This is remarkable because the SFR-normalized mass function of intermediate-age clusters can be affected by variations in the past formation and disruption rates of the clusters. The similarity of these CMFs indicates that the SFRs were relatively steady for the past 400~Myr, and that the cluster disruption rates were similar
among the galaxies in our samples.

In order to quantify the observed scatter, we fit the CMF/SFR for each galaxy
by the function:
  \begin{equation}
dN/dM = A \times \mathrm{SFR} \times (M/10^4\,M_{\odot})^{-1.9}~~,
 \end{equation}
 where we have fixed the power-law exponent to $-1.9$ (the mean of all 
$\beta$ values) for all galaxies. We use our H$\alpha$-based SFRs from Table~1, adjust them for partial coverage (if needed),  then normalize the distributions at $10^4\,M_{\odot}$, which is about the middle of the cluster log~mass range observed across the galaxies. The coefficient $A$ measures the proportionality between the cluster and star formation rates.  The dispersion $\sigma(\log\,A)$ in the best fit values of $\log\,A$,  quantifies the scatter in the amplitudes and hence in the CMF/SFR relation among galaxies.
 
We demonstrate our technique in Figure~3. The SFR-normalized cluster mass functions are shown in two intervals of age:
$\tau < 10$~Myr (left panel) and $\tau = 100$--400~Myr (right panel). The lines show the best fit from equation~(4), and the intercept values $A$ are the values where each fit intersects the dotted line. For the seven galaxies studied here, we find that $\sigma(\log\,A)=0.21$ for clusters with ages 
$\tau < 10$~Myr. The scatter for intermediate-age (100--400~Myr) clusters 
is similar, with $\sigma(\log\,A)=0.20$. This scatter is close to the dispersion expected from  random uncertainties in the CMFs and the SFRs,  as discussed in Sections~2.3 and 3.1.

The data presented here are consistent with exact proportionality between the formation rates of stars and clusters. Nevertheless, within the uncertainties, the data also allow for weak deviations, at the approximately factor of two level. In Figure~4, we search for trends in the residuals of $A$ with host galaxy properties, including SFR, luminosity, color, inclination,
maximum rotational velocity of the gas, and the exponent of the cluster mass function $\beta$. The nonparametric Spearman correlation coefficients, $r_S$
and the corresponding probabilities p($>r_S$) associated with each plot are compiled in Table~3. Values for p($>r_S$) should be less than 0.05 to have
greater than 95\% confidence that a correlation is real. For the young 
($\tau < 10$~Myr) clusters, we do not observe any statistically significant correlations. For the intermediate-age (100--400~Myr old) clusters, there are weak positive trends between the residuals of $\log\,A$ and galactic SFR, luminosity, and $B\!-\!V$ color, but only the one with $B\!-\!V$ is statistically significant (at the 95\% level). Possibly, these trends may reflect
differences in the metallicities of the galaxies. In any case, a larger sample of galaxies is needed to determine whether or not such trends are real (e.g., the LEGUS project; Calzetti et~al.\ 2015). We find no statistically significant correlation for clusters in either age range between the residuals of $\log\,A$
and the SFR per unit area or SFR density (not shown), which is often used to compare cluster populations among galaxies.

\subsection{Comparison with Previous Work}

The CMF/SFR statistic studied here is closely related to the cluster formation efficiency $\Gamma$, defined as the fraction of stellar mass formed in bound clusters. Estimates of $\Gamma$ in numerous studies have been used to draw important conclusions  about the properties of star-forming molecular clouds in different galaxies and environments. Since, as discussed in the Introduction, it is not possible to distinguish bound from unbound clusters based on images alone, $\Gamma$ should be directly proportional to our CMF/SFR statistic.

$\Gamma$ has been estimated for a number of galaxies, using various methods, and large differences are claimed from one galaxy to another.
All of these studies integrate over the CMF in order to determine the  total mass in clusters, but use different assumptions, extrapolations, age and mass ranges, and methods to accomplish this (e.g., Goddard et~al.\ 2010; Adamo et~al.\ 2011; Silva-Villa \& Larsen 2011; Silva-Villa et~al.\ 2013). The compiled results (e.g., Kruijssen 2012) show a strong increase in $\Gamma$,
by a factor of $\sim$15, for galaxies with increasing star formation rate densities, from $\approx$$10^{-3}$ to $\approx$ a~few~$M_{\odot}$~yr$^{-1}$~kpc$^{-2}$ (corresponding to the $\sim$3~dex range of SFRs from 
$\sim$0.05 to $\sim$50\,$M_{\odot}$ yr$^{-1}$). The corresponding relation between $\Gamma$ and SFR  (which has a similar slope when transformed from the original SFR density given in Kruijssen 2012) is shown as the dashed line in the top-left panel of Figure~4. This trend is much stronger than our CMF/SFR relation for young ($\tau < 10$~Myr) and
intermediate-age (100--400~Myr) clusters.

This may be due to the fact that estimates of $\Gamma$ are sensitive to the specific assumptions and extrapolations that are made,\footnote{In particular, 
in this approach the extrapolated number of clusters depends sensitively on the fit and uncertainties in the power-law index $\beta$ and  also on the exact mass and age ranges that are adopted. Small uncertainties in any of these values can drastically change the results.} and are thus prone to large errors. For example, Goddard et~al.\ (2010) found a significantly higher value of $\Gamma$ for clusters in the Antennae when compared with those
in the Magellanic Clouds, while we find similar CMF/SFR relations for these three galaxies. Our results are based on a larger cluster catalog  for the Antennae, but the same ones as used in Goddard et~al.\ (2010) for the LMC and SMC. The dotted line in the top-left panel of Figure~4 shows that the theoretical prediction for $\Gamma$ versus SFR from Kruijssen (2012) is much steeper than the observations, and therefore disagrees even more strongly with our results for the CMF/SFR statistic.

We have presented a robust new method for comparing the fraction of stars that form in clusters, one that only relies on comparing the amplitudes of the CMF normalized by the SFR in galaxies. The parameter $\Gamma$ has been extensively used in the literature, but requires additional assumptions
and extrapolations that degrade its reliability. We also advocate that, as done here, the SFR rather than the more uncertain SFR density be used when comparing results among galaxies. There is no standard method for estimating the star-forming area within galaxies, and we find that the uncertainties in area alone can be as large as a factor of a few (see also discussion in Adamo et~al.\ 2011).

\section{Conclusions and Implications}

The main conclusion of this paper is that star and cluster formation rates are proportional to each other in a diverse but representative sample of nearby galaxies. There are only minor deviations from exact proportionality in our
sample (with an rms scatter less than a factor of 2) in two different age ranges, and no trends are observed in the residuals of CMF/SFR with galaxy property
for young ($\tau < 10$~Myr) clusters. For the intermediate-age 
(100--400~Myr) clusters, weak trends with the star formation rate and luminosity of the host galaxy may be present, but it is also possible that
systematic uncertainties in the SFR determinations are responsible. A similar study with a larger sample is needed to establish whether or not weak trends exist.

Our results are consistent with previous findings that the luminosity of the brightest cluster in a galaxy is approximately proportional both to the number of clusters in the galaxy (e.g., Whitmore 2003) and to its star formation rate (Larsen 2002; Bastian 2008). However, inferences based only on the brightest cluster in a galaxy require extra (though plausible) assumptions about all the other, fainter clusters: their mass function, mass-to-light ratios, interstellar extinctions, and so forth. The method presented here is much more robust because it is based on the mass functions determined from hundreds or thousands of clusters over a wide range of masses and specific,
narrow ranges of age ($\tau < 10$~Myr and 100--400~Myr).

Our results are also consistent with our previous findings that the mass and age distributions of clusters are similar in different galaxies. We approximate the joint distribution by a double power law:
$g(M,\tau) \propto M^{\beta} \tau^{\gamma}$
with $\beta\approx -2$ and $-1.0 \lea \gamma \lea -0.5$ (Fall \& Chandar
2012, Chandar et~al.\ 2014). We refer to this as a quasi-universal distribution because the observed exponents differ relatively little among galaxies (typically, $\Delta \beta \sim \Delta \gamma \sim 0.2$).\footnote{
While the mass and age distributions of star clusters are remarkably similar on galaxy scales, in small regions there can be large variations in the formation and disruption rates, and hence in the observed mass and age distributions of the clusters (see e.g., Bastian et~al.\ 2012; Silva-Villa et~al.\  2014;  Chandar et~al.\ 2014 for an example of this effect in M83). As these regions are combined into larger ones, the variations will average out and the differences diminish. In particular, the exponent $\gamma$ of the age distribution may vary significantly with galactocentric distance for the physical
reasons discussed by Fall \& Chandar (2012).} Our results for the CMF/SFR relation imply that the populations of clusters on galaxy scales
are nearly the same apart from an overall normalization factor. The results presented here indicate that this normalization factor is the star formation rate.

Our claim that populations of clusters are similar in different galaxies might seem to be contradicted by the fact that massive clusters are found in some galaxies but not in others. For example, the most massive clusters in the Antennae have $M_{\mbox{max}}\sim10^7\,M_{\odot}$ whereas those in the SMC have $M_{\mbox{max}}\sim 10^4\,M_{\odot}$ (see Figure~1).
However, this is largely, if not entirely, due to the different sizes of these samples or populations, roughly $10^3$ times larger in the Antennae than in the SMC. Indeed, it is easy to show that, for the quasi-universal model, the maximum cluster mass scales with the population size and hence with the star formation rate as $M_{\mbox{max}} \propto SFR^{-1/(\beta + 1)}$, where 
$\beta$ again is the exponent of the mass function; thus $M_{\mbox{max}} \propto \mathrm{SFR}$ for $\beta=-2$.

The quasi-universal model implies that the clusters in different galaxies are formed and disrupted in similar (although not exactly the same) ways. This is a remarkable fact, probably the consequence of a great deal of similarity in the structure of the interstellar media among galaxies on the scale of molecular clumps and clouds, in particular in their mass functions, which play roles in both the formation and disruption of clusters (see Fall, Krumholz, \& Matzner 2010 and Fall \& Chandar 2012 for further discussion of these topics). Until recently, it has not been possible to observe the ISM of other galaxies with sufficient angular resolution to test this conjecture, but with the advent of ALMA, this may now be within reach.

\acknowledgments R.C.\ acknowledges support from NSF through CAREER
award 0847467.  S.M.F.\ acknowledges the hospitality of the Aspen Center for Physics, which is supported in part by the National Science Foundation under grant PHYS-1066293.  We thank the anonymous referee and Mark Krumholz for suggestions that improved our paper.

{\it Facilities:} \facility{HST}.

\begin{figure}
\plotone{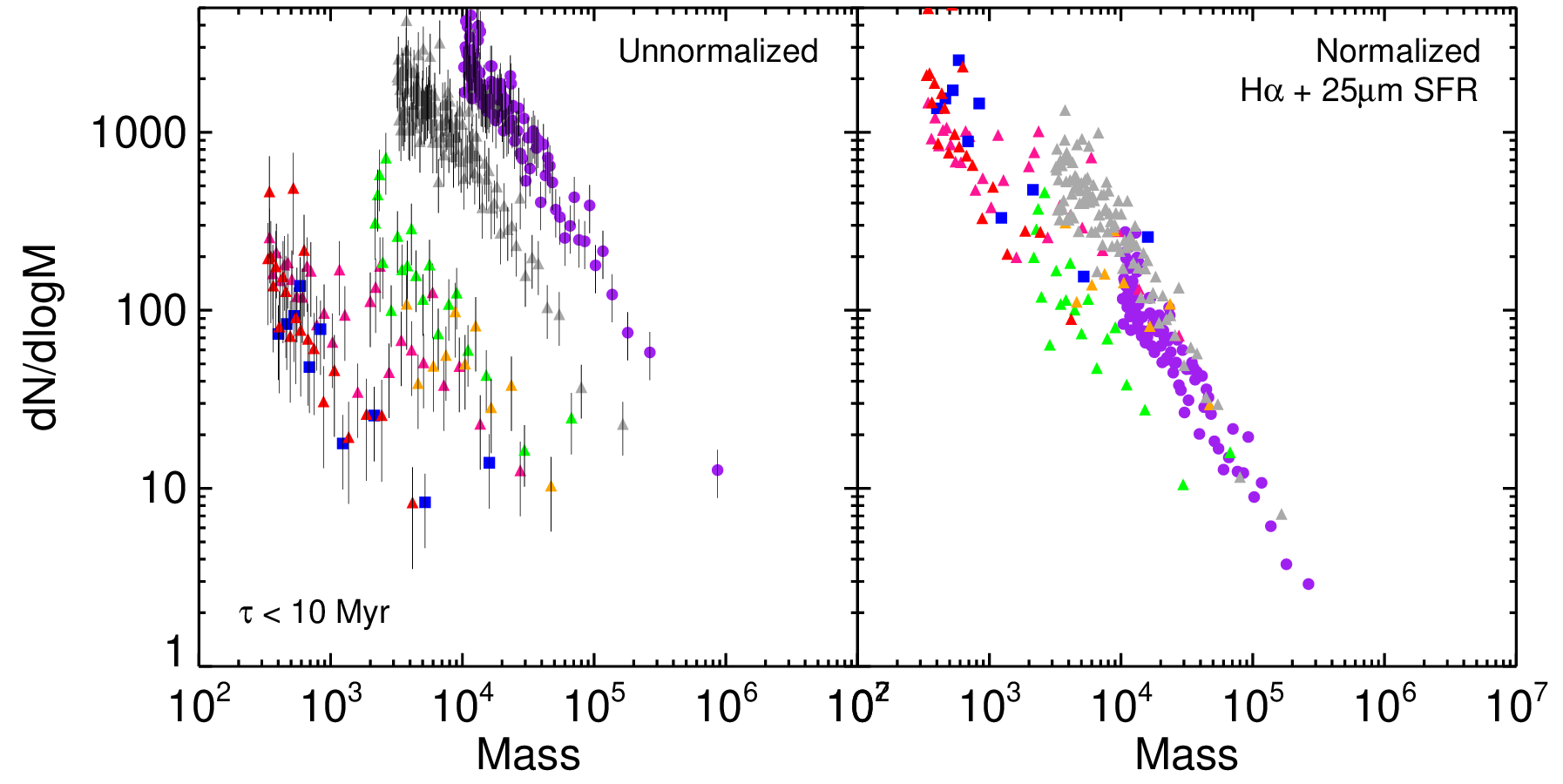}
\caption{Mass functions of star clusters younger than 10~Myr in 
7~star-forming galaxies.  The symbols are as follows: LMC: pink; SMC: blue; NGC~4214: red; NGC~4449: orange; M83: green; M51: gray; and Antennae: purple. The selection of the clusters is described in the text.  
The left panel shows the unnormalized mass functions.  The right panel shows the mass functions normalized by the current star formation rate (SFR).
The fractional coverage of each galaxy is also accounted for. }
\label{fig:mf}
\end{figure}

\begin{figure}
\plotone{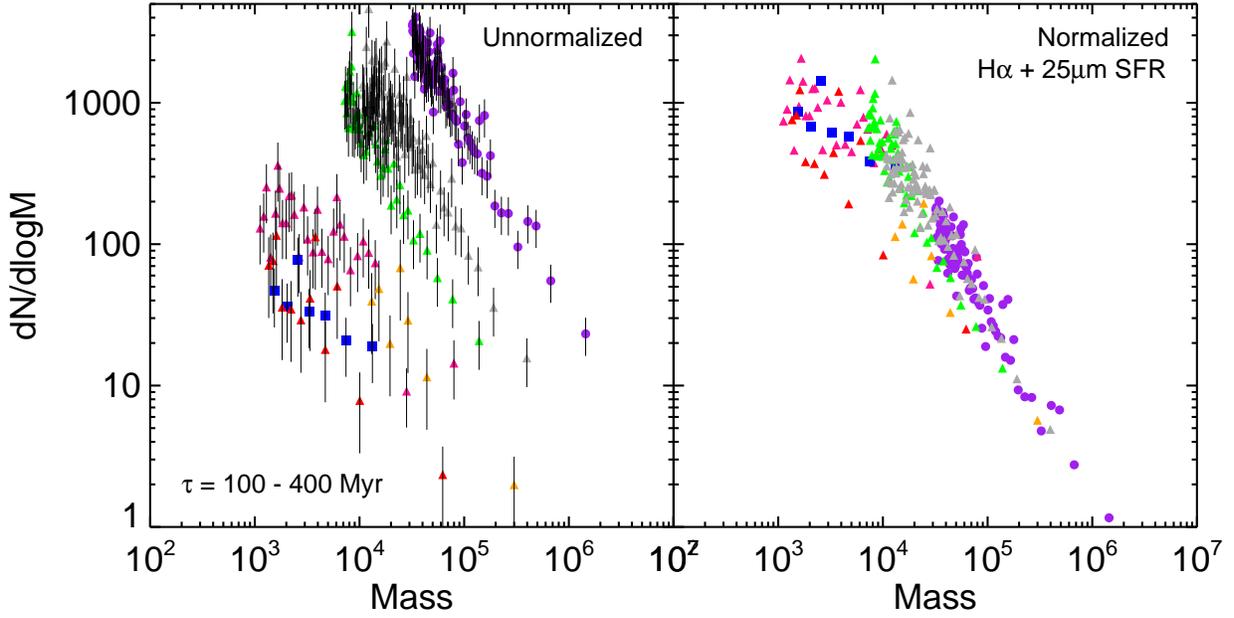}
\caption{Same as Figure~1, but now for clusters in the age range 100--400~Myr.}
\label{fig:mf}
\end{figure}

\begin{figure}
\plotone{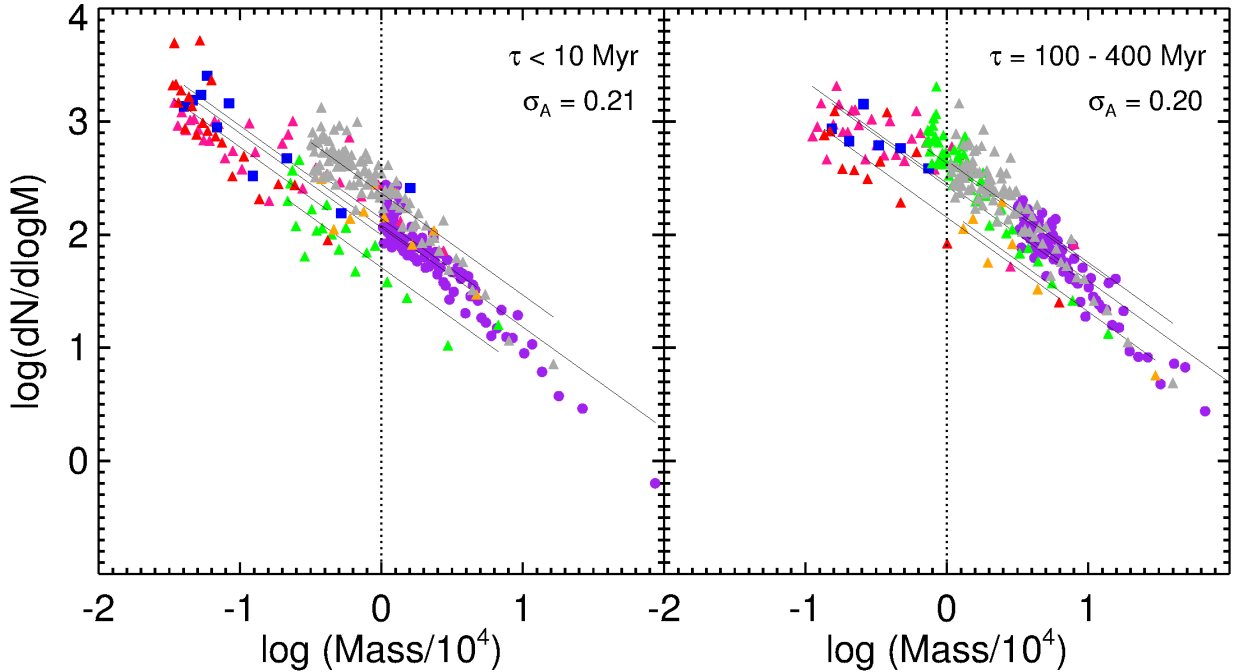}
\caption{Illustration of our method for determining the scatter in the CMF/SFR relation for young clusters ($\tau < 10$~Myr, left panel) and intermediate-age clusters ($\tau=100$--400~Myr, right panel). The standard deviations of $\log A$ are given in the upper right of each panel (denoted by $\sigma_A$).
}
\label{fig:sigmaA}
\end{figure}

\begin{figure}
\plotone{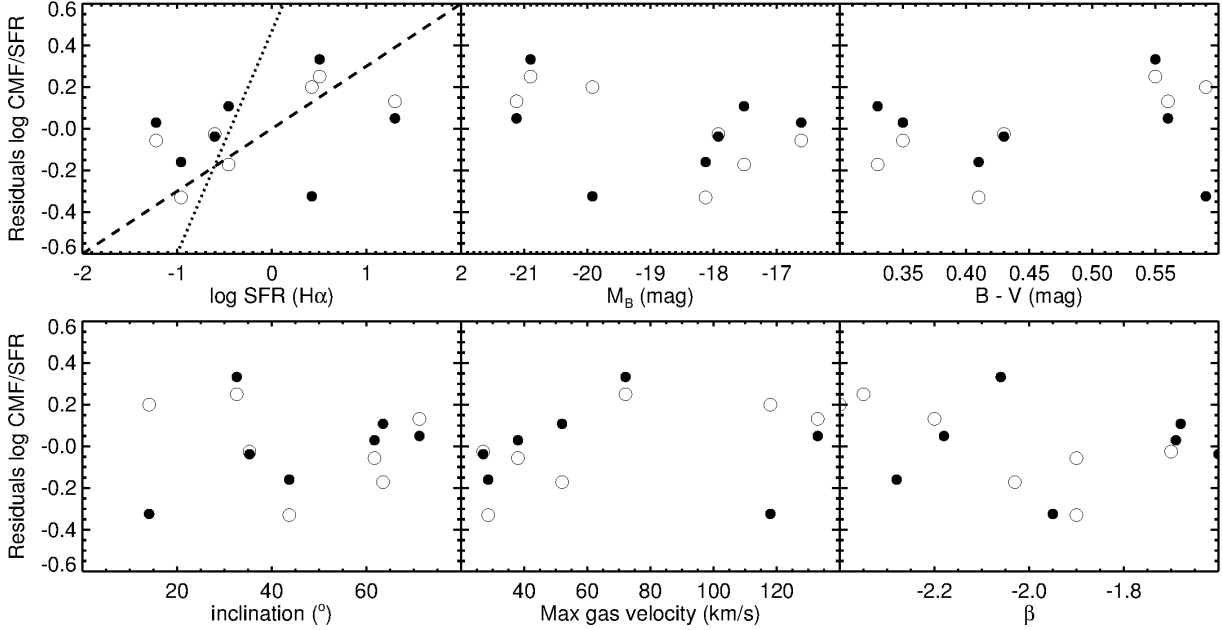}
\caption{Logarithmic residuals in the CMF/SFR relations for young 
($\tau \leq 10$~Myr, filled circles) and intermediate-age (100--400~Myr, open circles)  clusters plotted against a number of galaxy properties. No statistically significant correlations are found  between the residuals and the galaxy properties considered here for the young $\tau < 10$~Myr clusters.
There are weak trends with galactic SFR, luminosity, and $B\!-\!V$ color for the intermediate-age (100--400~Myr) clusters, but only the one with 
$B\!-\!V$ is statistically significant (at the 95\% level). The top-left panel shows that our results do not agree with the two correlations presented in
Kruijssen 2012, which were based on a compilation of observational results (dashed line) and theoretical calculations (dotted line). See text for more details. }
\label{fig:residuals}
\end{figure}

\clearpage
\begin{landscape}
\begin{deluxetable}{cccccccl}
\tablecolumns{3}
\tablecaption{Properties of Sample Galaxies\label{galaxies}}
\tablewidth{0pt}
\tablehead{
\colhead{}  & \colhead{} & \colhead{}  & \colhead{}  & \colhead{H$\alpha$ + 25$\mu$m} & \colhead{Published} & \colhead{Published} & \colhead{SFR} \\
\colhead{Galaxy}  & \colhead{Galaxy} & \colhead{Distance}  & \colhead{$\beta$\tablenotemark{1}} & \colhead{SFR\tablenotemark{2}} & \colhead{H$\alpha$ SFR} & \colhead{Other SFR} & \colhead{References} \\
\colhead{Name} & \colhead{Type} & \colhead{(Mpc)}  & \colhead{$\tau \leq 10$~Myr} &\colhead{($M_{\odot}/$yr)}  & \colhead{($M_{\odot}/$yr)}  & \colhead{($M_{\odot}/$yr)} & \colhead{}
}
\startdata
LMC  & Irregular &  0.0\rlap{5}   & $-1.60\pm0.10$  &   0.25  & 0.25 &  0.39 & 1, 2 \\
SMC &  Irregular & 0.0\rlap{6}  &  $-1.69\pm0.15$ &  0.06 & 0.04 &  0.05 & 3, 4   \\
NGC 4214 & Irregular &  3.1 & $-2.28\pm0.17$ &  0.11  & 0.16 &  0.22 & 5, 5 \\
NGC 4449 &  Irregular & 3.8  & $-1.68\pm0.21$  &  0.35 & 0.66 & 0.89 & 5, 5 \\
M83 &   Spiral & 4.5 & $-1.95\pm0.12$ &   2.65 & 3.35 & 5.01 & 5, 5  \\
M51 & Spiral & 8.2 & $-2.06\pm0.05$ & 3.20 & 4.48  & 7.58 & 5, 5 \\
Antennae & Merging & 22.2  & $-2.18\pm0.04$ & 20.20 & 25.48 & ... & 6  
\enddata
\tablenotetext{1}{Power-law index for the mass function of star clusters, determined from the least-squares fits to $\log(dN/d\log M) = (\beta-1)\log{M} + {\rm const}$}
\tablenotetext{2}{Star formation rates determined in this work.  See Section~3.1 for a description}
\tablecomments{ References:
1.~James et~al.\ 2008; 2.~Harris \& Zaritsky 2009; 3.~Bolatto et~al.\ 2011;  4.~Harris \& Zaritsky 2004; 5.~Lee et~al.\ 2009; 6. Zhang et~al. 2001
}                                                           
\end{deluxetable}
\end{landscape}

\clearpage
\begin{deluxetable}{cccccccl}
\tablecolumns{2}
\tablecaption{Uncertainties in SFR Estimates\label{tab:errors}}
\tablewidth{0pt}
\tablehead{
\colhead{Source} && \colhead{Uncertainty} \\
\colhead{of Uncertainty} && \colhead{log~SFR} 
}
\startdata
Cluster Mass Function  & & 0.1 ($\tau < 10$~Myr)\\
  && 0.04 ($\tau=100$--400~Myr) \\
Flux Measurements && $\sim$0.04--0.10 \\
Extinction Correction && $\approx$0.1 \\
Fractional Coverage && $\approx$0.0\rlap{4} \\
Lyc Escape Fraction && unknown\\
Metallicity && systematic trend \\
   &&  $\approx$0.4 at $1/10\,Z_{\odot}$ \\
\enddata
\end{deluxetable}


\begin{deluxetable}{cccccccl}
\tablecolumns{5}
\tablecaption{Spearman Correlation Coefficients for Residuals Shown in Figure~4\label{tab:fits}}
\tablewidth{0pt}
\tablehead{
\colhead{Galaxy} & \multicolumn{2}{c}{Age Range $\tau < 10$~Myr} & \multicolumn{2}{c}{Age Range $\tau=100$--400~Myr} \\
\colhead{Property} & \colhead{r$_S$} & \colhead{p($>\mbox{r}_S$)}  & \colhead{r$_S$} & \colhead{p($>\mbox{r}_S$)} 
}
\startdata
log~SFR ($M_{\odot}/$yr) & 0.39 & 0.39 & 0.71 & 0.07 \\
M$_{B}$ (mag) & \llap{$-$}0.11 & 0.82 & \llap{$-$}0.66 & 0.10 \\
$B\!-\!V$ (mag) & \llap{$-$}0.29 & 0.53 & 0.78 & 0.04 \\
inclination ($\deg$) & 0.39 & 0.38 & \llap{$-$}0.50 &0.25 \\
Max gas velocity (km/s) & 0.25 & 0.59 & 0.61 & 0.15 \\
CMF index $\beta$ & 0.07 & 0.88 & \llap{$-$}0.04 & 0.94  \\
\enddata
\tablecomments{The nonparametric Spearman correlation coefficient, $r_S$ and the corresponding probabilities p$(>r_S)$ are compiled.  A perfect correlation would have a value of $\pm 1$, and the sign indicates the direction of the association between the listed galaxy properties and the residuals in $\log\,A$. Values for p$(>r_S)$ should  be less than 0.05 to have greater than 95\% confidence that there is a real correlation.  The only correlation that satisfies this condition is the one with $B\!-\!V$ for intermediate-age clusters.} 
\end{deluxetable}


\begin{thebibliography}{}

\bibitem{ref1}
Adamo, A., Ostlin, G., \& Zackrisson, E. 2011, MNRAS, 417, 1904

\bibitem{ref1}
Annibali, F., Aloisi, A., Mack, J., et al. 2008, AJ, 135, 1900 

\bibitem{ref1}
Bastian, N. 2008 MNRAS, 390, 759
  
\bibitem{ref1}
Bastian, N., Adamo, A., Gieles, M., Silva-Villa, E., Lamers, 
H.~J.~G.~L.~M., Larsen, S.~S., Smith, L.~J., Konstantopoulos, I.~S., \&  Zackrisson, E. 2012, MNRAS, 419, 2606

\bibitem{ref1}
Bolatto, A.~D., Leroy, A.~K., Jameson, K., et al. 2011, ApJ, 741, 12

\bibitem{ref1}
Bruzual, G., \& Charlot, S. 2003, MNRAS, 344, 1000

\bibitem{ref1}
Calzetti, D., Lee, J.~C., Sabbi, E., et al. 2015, AJ, 149, 51

\bibitem{ref1}
Chabrier, G. 2003, PASP, 115, 763

\bibitem{ref1}
Chandar, R., Whitmore, B.~C., Calzetti, D., \& O'Connell, R. 2014, ApJ, 787, 17

\bibitem{ref1}
Chandar, R., Fall, S.~M., \& Whitmore, B.~C. 2010a, ApJ, 711, 1263 

\bibitem{ref1}
Chandar, R., Whitmore, B.~C., \& Fall, S.~M. 2010b, ApJ, 713, 1343 

\bibitem{ref1}
Chandar, R., Whitmore, B.~C., Kim, H., et al. 2010c, ApJ, 719, 966 

\bibitem{ref1}
Dopita, M.~A., Calzetti, D., Ma\'{\i}z Apell\'aniz, J., et al. 2010, Ap\& SS, 33

\bibitem{ref1}
Fall, S.~M., Krumholz, M.~R., \& Matzner, C.~D. 2010, ApJ, 710, L142

\bibitem{ref1}
Fall, S.~M., \& Chandar, R. 2012, ApJ, 752, 96

\bibitem{ref1}
Fitzpatrick, E.~L. 1999, PASP, 111, 63

\bibitem{ref1}
Fouesneau, M., Lancon, A., Chandar, R., \& Whitmore, B.~C. 2012, ApJ, 750, 60

\bibitem{ref1}
Goddard, Q.~E., Bastian, N., \& Kennicutt, R.~C. 2010, MNRAS, 405, 857

\bibitem{ref1}
Harris, J., \& Zaritsky, D. 2004, AJ, 127, 1531

\bibitem{ref1}
Harris, J., \& Zaritsky, D. 2009, AJ, 138, 1243

\bibitem{ref1}
Hunter, D., Elmegreen, B.~G., Dupuy, T. J., \& Mortonson, M. 2003, AJ, 126, 1836

\bibitem{ref1}
James, P.~A., O'Neill, J., \& Shane, N.~S. 2008, A\& A, 486, 131

\bibitem{ref1}
Kennicutt, R.~C. 1998, ARA\& A, 36, 189

\bibitem{ref1}
Kennicutt, R.~C., \& Evans, N.~J. 2012, ARA\& A, 50, 531

\bibitem{ref1}
Kruijssen, J.~M.~D. 2012, MNRAS, 426, 3008

\bibitem{ref1}
Lee, J.~C., Gil de Paz, A., Tremonti, C., et al. 2009, ApJ, 706, 599

\bibitem{ref1}
Leitherer, C., Schaerer, D., Goldader, J.~D., et al. 1999, ApJS, 123, 3

\bibitem{ref1}
Lada, C.~J., \& Lada, E.~A. 2003, ARA\&A, 41, 57 

\bibitem{ref1}
Larsen, S.~S. 2002, AJ, 124, 1393

\bibitem{ref1}
McKee, C.~F., \& Ostriker, E.~C. 2007, ARA\&A, 45, 565 

\bibitem{ref1}
Raiter, A., Schaerer, D., \& Fosbury, R.~A.~E. 2010, A\& A, 523, 64

\bibitem{ref1}
Rangelov, B., Prestwich, A.~H., \& Chandar, R. 2011, ApJ, 741, 86

\bibitem{ref1}
Salpeter, E. 1955, ApJ, 121, 161 

\bibitem{ref1}
Schweizer, F., Burns, C.~R., Madore, B.~F., et al. 2008, AJ, 136, 1482

\bibitem{ref1}
Silva-Villa, E., \& Larsen, S.~S. 2010, A\&A, 516, 10

\bibitem{ref1}
Silva-Villa, E., Adamo, A., \& Bastian, N. 2013, MNRAS, 436, 69

\bibitem{ref1}
Silva-Villa, E., Adamo, A., Bastian, N., Fouesneau, M., \& Zackrisson, E. 2014, MNRAS, 440, 116

\bibitem{ref1}
Smith, L.~J., Norris, R.~P.~F., \& Crowther, P.~A. 2002, MNRAS, 337, 1309

\bibitem{ref1}
Thim, F., Tammann, G.~A., Saha, A., et al. 2003, ApJ, 590, 256 

\bibitem{ref1}
Whitmore, B.~C. 2003, ``The Formation of Star Clusters,'' in A Decade of Hubble Space Telescope Science. ed., Mario Livio, Keith Noll, \& Massimo Stiavelli (Cambridge University Press, Cambridge, UK)

\bibitem{ref1}
Whitmore, B.~C., Chandar, R., \& Kin, H., et al. 2011, ApJ, 729, 78

\bibitem{ref1}
Zhang, Q., Fall, S.~M., \& Whitmore, B.~C. 2001, ApJ, 561, 727

\end{thebibliography}
\end{document}